\documentclass[12pt]{article}

\usepackage[utf8]{inputenc}
\usepackage[margin=1in]{geometry}
\usepackage{graphicx}
\usepackage{booktabs}
\usepackage{amsmath,amssymb}
\usepackage{natbib}               
\usepackage{hyperref}
\usepackage{subcaption}
\usepackage{tabularx}

\setcitestyle{authoryear,open={(},close={)}}

\title{\textbf{See You at the Posterior Line: Learning Bayesian Modeling Through a Car Racing Game}}

\author{
  Federica Zoe Ricci$^{1,*}$ \quad Mine Dogucu$^{2}$
}

\date{
  \small
  $^{1}$Department of Mathematics and Statistics, Swarthmore College, Swarthmore, USA\\
  $^{2}$Department of Statistics, Harvard University, Cambridge, USA\\[2ex]
  $^{*}$\textbf{Corresponding Author:} fricci1@swarthmore.edu\\[1ex]
  \textbf{ORCID iDs:}\\
  Author One: \url{https://orcid.org/0000-0002-2560-4543}\\
  Author Two: \url{https://orcid.org/0000-0002-8007-934X}
}

\begin{document}
\maketitle

\begin{abstract}
We present an interactive classroom activity designed to address a central challenge in teaching introductory Bayesian statistics: how to formalize subjective knowledge and available information into prior distributions and then update them with empirical data. Role-playing as data analysts for a racing team, students evaluate candidate tires by converting qualitative engineering reports into prior distributions, collecting primary data via a virtual racing game, and using a Beta-Binomial model to inform team strategy. This discovery-based exercise allows small groups to observe directly how different prior choices and sample data jointly shape posterior inference. Student feedback ($n=32$) highlights high enjoyment, engagement and improved conceptual clarity. Open-access materials to implement the activity are provided, alongside recommendations for adapting it to other teaching contexts.

\end{abstract}

\section{Introduction}

Bayesian statistics frames learning from data as a sequential process: an analyst's initial knowledge and uncertainty about a parameter are formalized into a prior distribution, which is then updated as new data are observed. Familiarizing students with this paradigm is a core objective of introductory Bayesian courses. However, because Bayesian statistics is typically taught after students have already been trained in frequentist methods \citep{dogucu2022current}, this shift in perspective presents a major conceptual challenge.

In particular, understanding what prior distributions represent and how they arise from real-world information is a central hurdle for novice Bayesian learners \citep{hu2022content}. When first encountering the Bayesian paradigm, students frequently struggle with fundamental questions: \textit{What are concrete examples of prior information? How does qualitative information translate into mathematical parameters? Is there only one correct prior?} 
Providing satisfying, hands-on answers in the classroom remains difficult, because traditional textbook examples often presents priors as given, rarely focusing on the practice of prior specification itself. Beyond prior construction, there is a broader scarcity of interactive resources that let students experience the full sequential workflow—from eliciting a prior distribution, to gathering data and then observing how different, reasonable prior choices shape posterior inference.

Integrating game-based simulations into statistics courses stands as a promising solution to this pedagogical challenge \citep{cummiskey2012using, kuiper2015using}. Generating primary data locally through interactive games fosters student ownership \citep{kuiper2025greenhouse} and reduces learning anxiety \citep{lesser2008functional}. Crucially for Bayesian education, game-based activities challenge the notion that statistical studies always have a single ``correct'' answer \citep{kuiper2015using}. By combining qualitative scenario information with student-generated data, instructors can create a realistic decision-making environment where multiple valid prior specifications naturally emerge and converge through empirical observation.

We present \textit{See You at the Posterior Line}, an interactive classroom activity designed to guide students through the complete Bayesian workflow—from prior construction to posterior interpretation. 
Intended for use when students are first introduced to Bayesian modeling, the activity assumes only an initial exposure to the Beta-Binomial model and access to a tool to plot Beta distributions.
The exercise places students into small teams role-playing as data analysts for a racing team. In the sections that follow, we outline the teaching context, detail the components of the activity and describe classroom management, present student evaluation feedback ($n=32$), and offer practical recommendations for adapting the activity to other instructional settings.

\section{Teaching Context and Implementation}

The activity was implemented early in the term during the Beta-Binomial unit of an undergraduate \textit{Introduction to Bayesian Analysis} course at a large research university. Because the exercise requires only a foundational understanding of Beta-Binomial updating, it can easily be integrated into any introductory statistics or data science module that introduces Bayesian principles. In our specific context, the 10-week, in-person course enrolled 50 students, primarily Data Science majors with prior probability coursework and basic familiarity with R. The broader course progression moves from conjugate models (Beta-Binomial, Gamma-Poisson, and Normal-Normal) to the basics of Markov Chain Monte Carlo, posterior inference, and regression models.

Delivered during a 50-minute class session, the activity was completed in small groups of 4 to 5 students. Each student requires an internet-connected computer to play the online racing game. In addition, the activity requires access to a tool capable of plotting Beta distributions, which is essential for students to carry out their prior choices and to visualize and interpret the resulting posteriors. In our implementation, we provided students with a scaffolded Quarto worksheet in RStudio. Because students were already familiar with R, we used functions from the \textbf{bayesrules} package \citep{dogucu2022bayesrules} to streamline Beta-Binomial visualization and updating. However, the activity is software-agnostic: any tool capable of plotting Beta distributions—such as interactive web applets, spreadsheets, or Python—can easily be substituted.

The activity employs a guided discovery approach using a worksheet that prompts students when to deliberate in their groups, what key information and decisions to record, and when to work individually. Consequently, the instructor's role is primarily supportive: introducing the scenario, facilitating group formation, help managing time across activity phases, and addressing occasional questions. While this initial session was co-facilitated by the activity developer and a graduate teaching assistant, a single facilitator can comfortably manage a class of 50 students. For larger class sizes, additional facilitators may be helpful depending on students' need for technical troubleshooting with the chosen plotting software. Although delivered in person, the activity structure can translate to synchronous online settings using breakout rooms.

Data for evaluating the activity was collected under an Institutional Review Board (IRB) exemption protocol. All students present during the class session participated in the activity as standard coursework, and completed a brief post-lab reflection survey assigned as homework. To ensure ethical compliance and eliminate potential bias, student consent for research participation was gathered via a separate form. The primary course instructor remained blinded to consent decisions until final grades were submitted to the registrar, and only de-identified responses from consenting students ($n=32$) were subsequently analyzed and reported.

\section{The Racer Game}
\label{sec:racer}

To provide a time-effective data-generating mechanism, our activity utilizes \textit{Racer} (\url{https://www.stat2games.sites.grinnell.edu/games/raceradvanced22.html}), an online game hosted on the free Stat2Games platform (\url{https://www.stat2games.sites.grinnell.edu/}). Designed specifically for statistics and data science education, Stat2Games provides browser-based game applets that allow students to generate primary datasets through interactive play without requiring user accounts or software installation \citep{kuiper2015using}.

\begin{figure}[htbp]
    \centering
    \begin{subfigure}{0.45\linewidth}
        \centering
        \includegraphics[width=\linewidth]{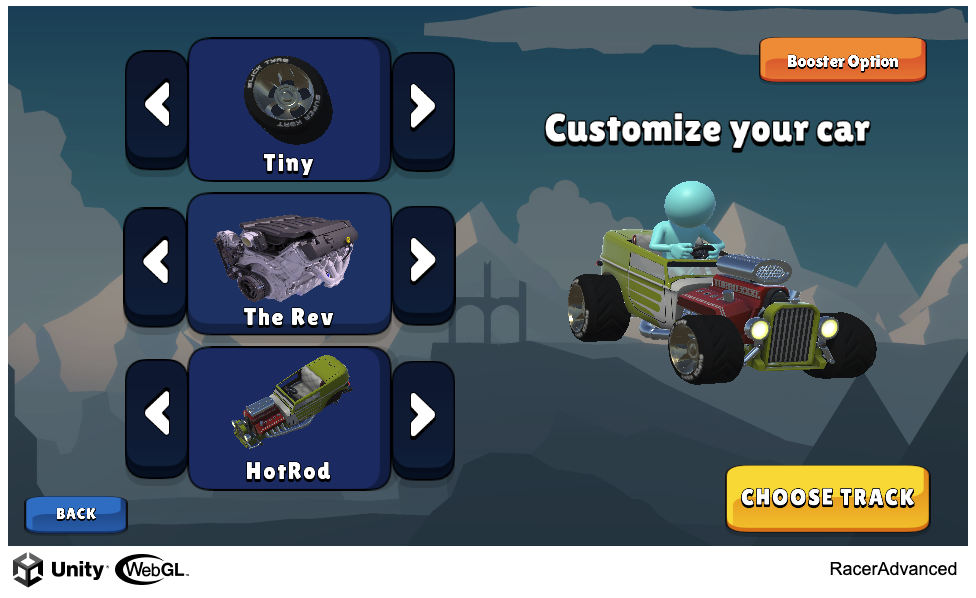}
        \caption{Customization menu with tire,  engine and car body selection options.}
        \label{fig:racing-game-a}
    \end{subfigure}
    \vspace{0.5em}
    \begin{subfigure}{0.45\linewidth}
        \centering
        \includegraphics[width=\linewidth]{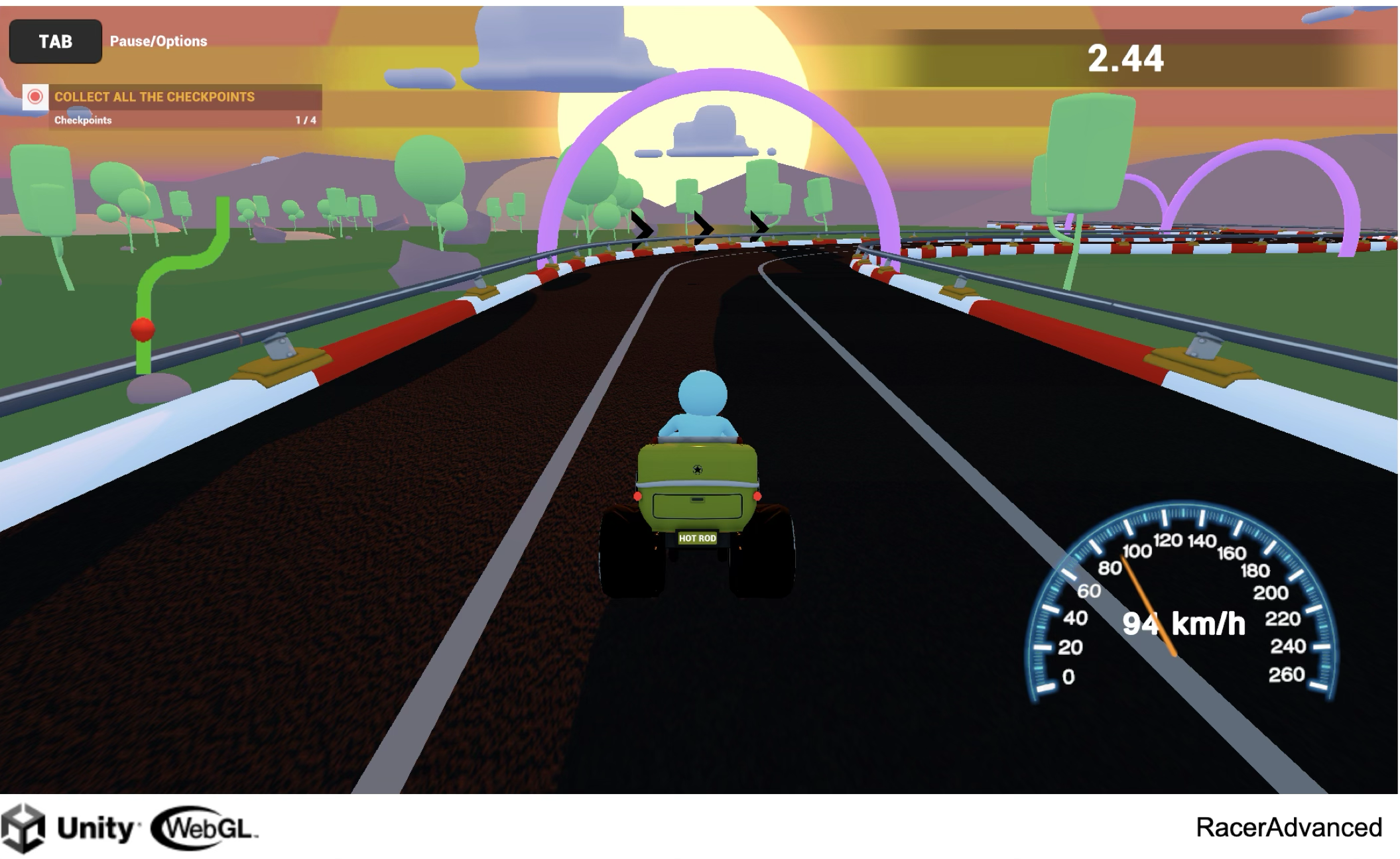}
        \caption{Real-time gameplay interface on the "Eight" track in the \textit{Racer} applet.}
        \label{fig:racing-game-b}
    \end{subfigure}
    \caption{Interface overview of the \textit{Racer} game on Stat2Games: (a) configuring tire selection in the car customization menu, and (b) driving a trial lap on the "Eight" circuit.}
    \label{fig:racing-game}
\end{figure}

When launching \textit{Racer}, players enter arbitrary Player and Team IDs; these can be used to retrieve data from a specific race  upon downloading data from Stat2Games, which we do not require for this activity. After entering some IDs, students  select the car customization menu. Then, to isolate tire performance, they select the ``HotRod" car body with default engine settings, equipping either ``Tiny" or ``HotRod" tires  (see Figure~\ref{fig:racing-game}, Left), as instructed by the activity worksheet. Players then select the ``Eight" track—a figure-eight circuit requiring approximately 1 minute to play using standard keyboard arrow controls (see Figure~\ref{fig:racing-game}, Right).

The track and tire parameters were deliberately chosen to create an accessible yet authentic, non-trivial game experience. The ``Eight" track is  a balanced circuit, complex enough to engage students with gaming experience, without being so difficult as to frustrate non-gamers.  Pre-testing revealed that ``Tiny" and ``HotRod" tires yield very similar average lap times on this track, ensuring meaningful variation in student-collected lap data, preventing a trivially obvious winner and creating a genuine modeling challenge.

\section{The Racing Scenarios}
\label{sec:racing-scenarios}
\begin{figure}[htbp]
    \centering
    \includegraphics[width=0.6\linewidth]{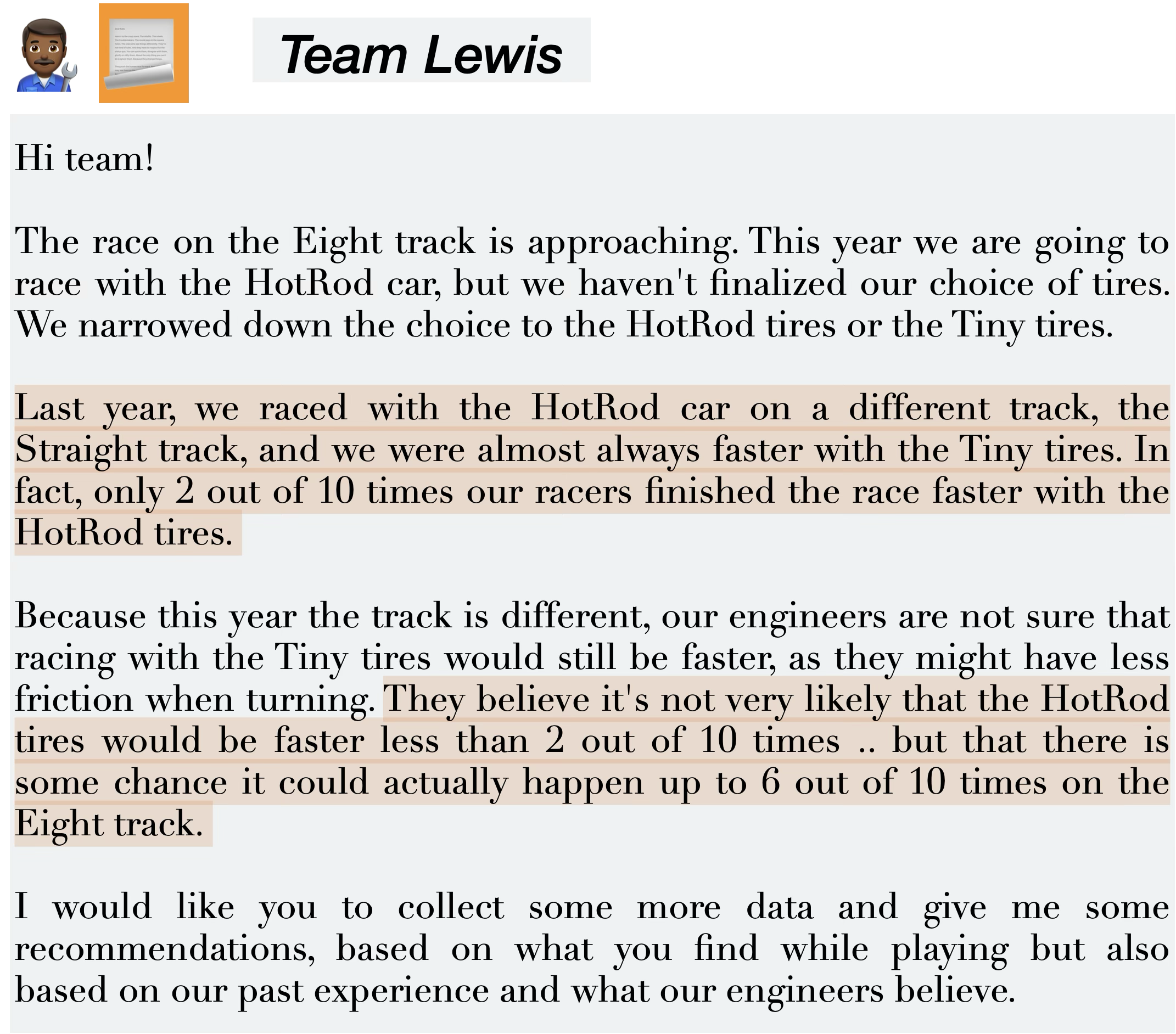}
    \caption{Scenario letter from Lewis Hamilton's racing team. Qualitative background and engineering estimates highlighted in orange vary across the three scenario versions.}
    \label{fig:lewis-letter}
\end{figure}

To simulate a real-world Bayesian modeling context where domain experts hold varying pre-existing beliefs based on distinct yet equally valid viewpoints, student teams are assigned one of three scenario letters (Figure~\ref{fig:lewis-letter}). To foster thematic engagement, each letter is authored under the persona of a prominent motorsport figure—Danica Patrick, Lewis Hamilton, or Mario Andretti—requesting a data-driven recommendation on whether to equip ``HotRod'' or ``Tiny'' tires on the ``HotRod'' car for an upcoming race on the ``Eight'' track.

Each scenario combines qualitative domain context, past racing summaries, and engineering opinions. In the activity, teams will be prompted to translate this background into a Beta prior distribution $\text{Beta}(\alpha, \beta)$ for the parameter $\pi$, defined as the underlying probability that a driver finishes the race faster using HotRod tires:

\begin{itemize}
    \item \textbf{Danica Patrick's Team (Left-Skewed Prior):} Summarizes past race data on the Eight track (where HotRod tires were faster in 3 out of 4 trials), but notes caveats such as the use of a different car body in those trials. Engineers expect HotRod tires to win around 6 out of 10 races, with a plausible range between 4 and 9 out of 10.
    
    \item \textbf{Lewis Hamilton's Team (Right-Skewed Prior):} Shares historical data from a different track where HotRod tires performed poorly (faster in only 2 out of 10 trials). Accounting for track differences, engineers expect HotRod tires to be faster between 2 and 6 out of 10 times on the Eight track (Figure~\ref{fig:lewis-letter}).
    
    \item \textbf{Mario Andretti's Team (Symmetric Prior):} Represents a new team with no historical track data. Engineers have no reason to favor one tire over the other, believing a 2 out of 10 win rate is just as likely as an 8 out of 10 win rate, while explicitly ruling out extreme outcomes where one tire always leads to a better finishing time.
\end{itemize}

Distributing these three scenario variants across the class serves a central pedagogical goal: demonstrating why multiple, distinct prior specifications can be reasonable in applied practice. First, because historical observations are sparse or imperfect and must be synthesized with qualitative engineering estimates, even groups assigned the same scenario letter will encode slightly different prior parameters. Second, groups assigned different manager letters begin from fundamentally distinct baselines. During the class debrief, students evaluate the validity of these differing prior specifications and observe firsthand how initial beliefs are influenced or overridden as empirical game data is collected.

\section{Description of the Activity}

Designed for a 50-minute class session, the activity unfolds across five sequential phases: activity setup (5~min), three sections of the worksheet -prior choice (10~min), data collection (10~min), posterior analysis (10~min)-and a class-wide wrap-up discussion (15~min). The following sections detail the structure for each phase.

\subsection{Set-up: activity briefing and team preparation}

The instructor announces to students that they will play a racing game in small groups to practice applying the Beta-Binomial model introduced in prior coursework. Following a brief demonstration of the \textit{Racer} interface—highlighting car customization and track selection (see Section 2)—the instructor facilitates the formation of teams of four or five students.

Each team receives a physical envelope containing role assignment slips and their manager letter (see Section 3). Group members randomly draw one or more of five roles, with a description of the corresponding responsibility: \textit{Facilitator} (moderates discussions to ensure all members contribute to group discussions), \textit{Summarizer} (synthesizes consensus), \textit{Annotator} (records  consensus on the worksheet), \textit{Timer} (controls that the group moves along the activity phases following the recommended completion times), and \textit{Speaker} (asks potential clarifying questions to the instructor and may present team findings). Teams open their scaffolded worksheet and complete a brief ice-breaker: tell each other their names and assigned roles, and complete a lighthearted task (identify the most unrealistic among team members' New Year resolutions --e.g., Team 1  entered "Gain 10 pounds").

\subsection{Worksheet part 1: prior specification}

Teams review their manager's letter and translate the qualitative narrative into a formal Beta prior distribution, $\text{Beta}(\alpha, \beta)$. The worksheet guides students through four scaffolded prompts:
\begin{enumerate}
    \item Summarizing available historical data and engineering beliefs.
    \item Selecting a center for the prior probability distribution.
    \item Identifying strengths and pitfalls of the prior information.
    \item Interactively tuning $\alpha$ and $\beta$ hyperparameters using the \texttt{plot\_beta()} function of bayesrules, until a desired prior is found.
\end{enumerate}

\begin{figure}[htbp]
    \centering
    \includegraphics[width=\linewidth]{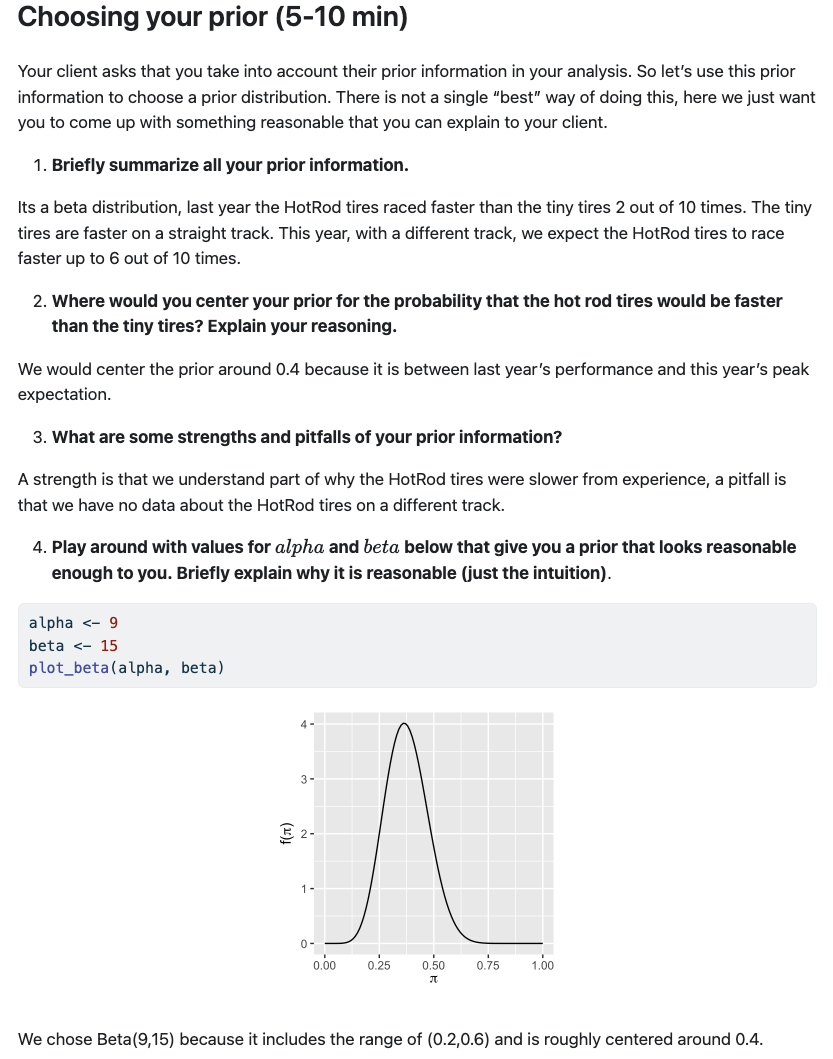}
    \caption{Scaffolded prior elicitation section of the Quarto worksheet, showing responses and  prior visualization from Team 1.}
    \label{fig:prior-choice}
\end{figure}

To emphasize that Bayesian prior elicitation involves reasonable modeling choices rather than a single ``correct" formula, students must explicitly justify their parameters. For instance, Team 1  (assigned Lewis Hamilton's scenario) noted that HotRod tires won only 2 out of 10 previous races on a straight track, but engineers expected up to 6 out of 10 on the Eight track. Balancing these inputs, they centered their prior near $0.40$ and selected a $\text{Beta}(9, 15)$ distribution (Figure~\ref{fig:prior-choice}).

\subsection{Worksheet part 2:  data collection via gameplay}

Next, teams generate primary experimental data on the Stat2Games platform. To familiarize themselves with game controls, each student first completes a warm-up  lap on the ``Eight" track using the baseline ``Classic'' car. Students then collect data by racing the ``HotRod'' car twice: once with HotRod tires and once with Tiny tires. To control for order and learning effects, tire sequence is counterbalanced within each group—the Facilitator and Annotator race first with HotRod tires, while the remaining team members race first with Tiny tires.

Students record their exact finishing times and determine individual binary outcomes (1 if HotRod was faster, 0 otherwise). Group members then aggregate their results in a pre-formatted code chunk to calculate the total successes $y$ out of $n$ trials. In Team 1, two of $n = 5$ members recorded faster lap times with HotRod tires ($y = 2$).

\subsection{Worksheet part 3: posterior updating and recommendation}

In the final worksheet phase, students execute pre-written R code using \texttt{summarize\_beta\_binomial()} from bayesrules and \texttt{plot\_beta\_binomial()} to compute and plot their posterior distribution, $\text{Beta}(\alpha + y, \beta + n - y)$. 

\begin{figure}[htbp]
    \centering
    \includegraphics[width=\linewidth]{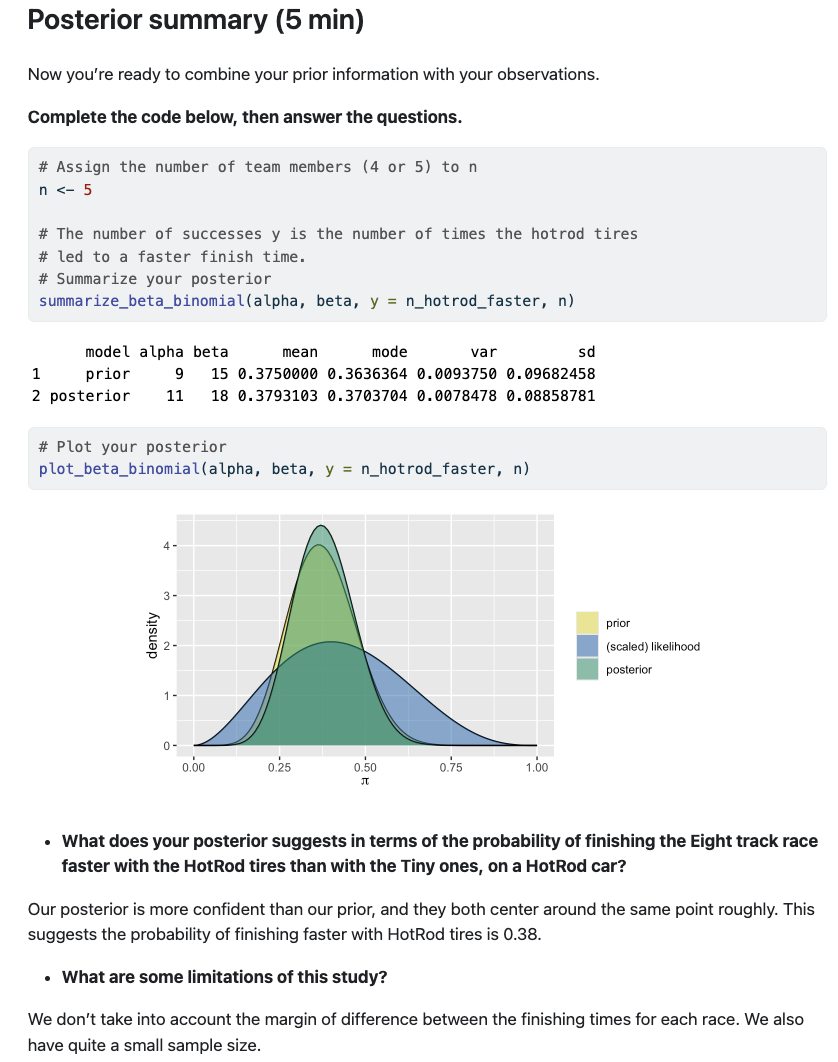}
    \caption{Posterior synthesis phase filled out by Team 1, showing numerical summary, distribution overlay, strategic recommendation, and critique of study limitations.}
    \label{fig:posterior}
\end{figure}

Teams synthesize their findings by writing a recommendation to their manager and assessing study limitations (Figure~\ref{fig:posterior}). Combining their $\text{Beta}(9, 15)$ prior with $y = 2$ successes across $n = 5$ trials, Team 1 updated to a $\text{Beta}(11, 18)$ posterior (posterior mean $\approx 0.38$). In their report, the team advised against using the HotRod tires, while noting critical methodological limits: the small sample size ($n = 5$) and the omission of margin-of-victory magnitude in a binary success metric.

\subsection{Wrap-up: class-wide debrief (10 min)}

To conclude the session, the instructor facilitates a class-wide discussion. Because Quarto worksheets render directly into clean HTML presentations, nominated \textit{Speakers} project their team's rendered document without additional preparation.

The instructor leverages differences across groups to reinforce key Bayesian concepts:
\begin{itemize}
    \item \textbf{Subjectivity in elicitation:} Comparing groups assigned the same scenario letter highlights how different teams encode identical qualitative text into distinct priors (e.g., $\text{Beta}(3, 5)$ vs. $\text{Beta}(9, 15)$), demonstrating the role of subjective judgment in hyperparameter choice.
    \item \textbf{Incorporating domain expertise:} Comparing outputs across different manager letters illustrates how distinct initial baselines shape posterior estimates., and exemplifies how Bayesian analysis allows to integrate qualitative expert knowledge into data-driven decision-making.
    \item \textbf{Asymptotic convergence:} Follow-up questions prompt students to consider what would happen if sample sizes increased dramatically ($n \to \infty$). Students recognize that as empirical evidence grows, the influence of initial prior specifications diminishes, causing disparate posterior distributions to converge.
\end{itemize}

\section{Student Feedback}

To evaluate student perceptions, an end-of-lab survey was administered featuring four open-ended prompts regarding: (1) positive learning impacts, (2) negative learning impacts, (3) overall impressions, and (4) suggestions for improvement. Of the 50 enrolled students, 42 participated in the lab session, and 36 consented to research use of their responses. After excluding four students who answered questions regarding a concurrent textbook homework rather than the lab activity, valid responses were obtained from 32 students ($n = 32$). Full anonymized survey responses are available in the Supplementary Materials.

Responses were analyzed jointly across all four prompts using inductive thematic coding to identify recurring themes regarding the activity's pedagogical strengths and logistical challenges.

\subsection{Activity's Strengths}

Nearly all respondents (31 of 32) cited at least one positive aspect of the activity, with most highlighting multiple features. Table~\ref{tab:student-feedback} summarizes the primary positive themes identified from the qualitative thematic analysis.

\begin{table}[htbp]
    \centering
    \caption{Primary activity's strengths identified in qualitative student feedback ($n = 32$). }
    \label{tab:student-feedback}
    \begin{tabularx}{\linewidth}{X c c X}
        \toprule
        \textbf{Theme} & \textbf{Count ($n$)} & \textbf{\%} & \textbf{Description} \\
        \midrule
        Fun and Enjoyment & 22 & 68.8\% & Described the lab as fun or enjoyable. \\
        Conceptual Understanding & 17 & 53.1\% & Reported improved grasp of Bayesian modeling workflows. \\
        Engagement, Interactive Format & 15 & 46.9\% & Valued the hands-on design over passive lectures. \\
        Team Collaboration & 14 & 43.8\% & Appreciated working in structured group roles. \\
        Real-World Context & 10 & 31.3\% & Noted the activity helped them appreciate the real-world applicability of course materials. \\
        Primary Data Collection & 9 & 28.1\% & Valued working with data they generated themselves. \\
        Gameplay & 9 & 28.1\% & Enjoyed the novelty of game-based learning. \\
        \bottomrule
        \multicolumn{4}{p{\linewidth}}{\small \textit{Note:} Percentages reflect unprompted thematic mentions in open-ended feedback. Omission of a theme by a respondent indicates it was not spontaneously cited, rather than a negative evaluation of that aspect.}
    \end{tabularx}
\end{table}

Representative qualitative responses illustrate the themes in Table \ref{tab:student-feedback}:
\begin{quote}
\textit{“The racing game was very fun which made learning the concepts much more engaging than simply reading about a simulation in a book.”}
\end{quote}
\begin{quote}
\textit{“Interactive, applicable to daily life, working in groups simulated working collaboratively in a workplace setting.”}
\end{quote}
\begin{quote}
\textit{“Being able to use R and see how our race results impacted the posterior model had a positive impact on my learning.”}
\end{quote}

A subset of respondents ($n = 10$) expressed exceptional enthusiasm for the format, with some of these explicitly pointing to long-term knowledge retention:
\begin{quote}
\textit{“It connected what we are learning to a practical and fun example that will help me remember the concepts better long term.”}
\end{quote}
\begin{quote}
\textit{“I genuinely loved it. It honestly is the most fun activity I have had in the past four years, but I was also able to learn a lot... Overall was a really great way to implement what we had been learning.”}
\end{quote}

\subsection{Constructive Critiques}

Constructive feedback focused primarily on logistical pacing and room setup:
\begin{itemize}
    \item \textbf{Time Constraints ($n = 18$):} Feeling rushed during the 50-minute class period, particularly during prior elicitation and gameplay.
    \item \textbf{Classroom Infrastructure ($n = 6$):} Communication difficulties caused by single-row seating arrangements.
    \item \textbf{Minor Implementation Challenges:} A small number of students noted minor confusion over written instructions ($n = 3$), felt group roles were unnecessary, or suggested smaller teams to ensure individual accountability. One student reported struggling to connect the gameplay directly to lecture material.
\end{itemize}

Strategies for addressing these implementation challenges are discussed in the concluding section.

\section{Discussion and Conclusion}

The primary aim of \textit{See You at the Posterior Line} is to improve student engagement with, and conceptual understanding of, Bayesian modeling through an authentic, narrative-driven context. Student evaluation of the activity was overwhelmingly positive: 31 of 32 respondents explicitly highlighted positive learning impacts, praising its fun, interactive nature, and noting that connecting practical play to R modeling improved both immediate conceptual understanding and long-term retention.

The design of this activity directly incorporates the GAISE recommendations on integrating real data within a meaningful, authentic context \cite{GAISE2016}, extending these principles into upper-level Bayesian instruction as advocated by \cite{hu2022content}. Following a discovery-based learning approach, students navigate prior elicitation and data collection from first principles. From a learning perspective, the process of group discussion and prior choice is at least as important as the resulting mathematical derivations.

The activity offers a classroom-scale glimpse into applied statistical consulting. Translating qualitative manager letters into quantitative Beta priors puts students in direct contact with the inherent ambiguity of prior elicitation, while demonstrating the practical utility of the Bayesian framework. During the class debrief, comparing groups assigned the same manager letter demonstrates how different teams encode identical qualitative information, while comparing across different letters illustrates how initial prior beliefs anchor posterior estimates in small-sample settings—and how posterior distributions eventually converge as empirical evidence accumulates.

Three practical points should be noted before implementing this exercise. First, instructors  should complete several trial laps on the racing game beforehand, to learn how to demonstrate the game to the class and anticipate questions regarding controls. Second, pacing within a 50-minute class period was a primary challenge, which led to compressing the final class-wide debrief. If possible, we recommend running the activity in a 75-minute block, or splitting it across two class meetings (e.g., prior elicitation, data collection and posterior interpretation in Session 1, class-wide debrief in Session 2). Third, physical classroom layout matters: several students noted that fixed row seating made group communication harder, so hosting the lab in a space with round or movable tables is advised, if possible.

Limitations of our evaluation should also be acknowledged. Evaluation measures were based on self-reported student feedback rather than objective pre- and post-tests of statistical literacy, and no randomized control group was used. Additionally, the exercise has only been tested in a face-to-face format; while the Quarto worksheet and Stat2Games platform could adapt to online breakout rooms, this delivery mode remains untested.

Finally, extensions are readily possible across different software, statistical models, and course formats. For instance, if students lack computing or programming experience, the exercise can be adapted into a paper-based worksheet paired with interactive web tools for Beta distribution visualization and Beta-Binomial updating—such as \textit{Probability Playground} \cite{probabilityplayground} or Brown University's \textit{Seeing Theory} visualizer \cite{seeingtheory}. Alternatively, for computational courses, R code chunks in the Quarto document can be replaced with Python syntax and mirrored in Jupyter Notebooks using libraries such as \texttt{PyMC} or \texttt{SciPy}. Because \textit{Racer} records continuous variables (e.g., exact lap times, drag, top speed), manager letters and worksheets can be adapted for Normal-Normal conjugate models or logistic regression. Instructors wishing to build similar activities around other Stat2Games applets should select tasks with non-obvious performance trade-offs to ensure meaningful data variability. Overall, as highlighted by the student survey, this activity offers an engaging, adaptable framework that bridges abstract Bayesian theory and practical data analysis.

All activity materials—including lesson slides, envelope scenario prompts, student worksheets, and sample solutions—are provided in the online supplementary repository (\url{https://anonymous.link}).

\section*{Supplementary Materials}

All files required to implement and reproduce this activity are available in the supplementary repository. The materials are organized into three primary sub-folders corresponding to the implementation workflow, along with an exemplar submission folder:

\begin{enumerate}
    \item \texttt{1-slides}: Contains editable Quarto (\texttt{.qmd}) and compiled PDF slides used to introduce the activity and guide students through the lab.
    \item \texttt{2-envelopes}: Contains print-ready files for physical envelope preparation:
    \begin{itemize}
        \item \texttt{group-names.md}: Team names and brief character bios to be printed and attached to envelope exteriors.
        \item \texttt{group-scenarios.md}: Qualitative manager letters to be printed and inserted inside envelopes.
        \item \texttt{group-roles.md}: Group member role definitions and descriptions.
        \item \texttt{envelopes-prepared.jpeg}: A reference photograph showing fully assembled envelopes.
    \end{itemize}
    \item \texttt{3-worksheet}: Contains \texttt{worksheet.qmd}, the interactive Quarto template provided to students. It guides teams through the activity workflow, incorporates embedded reflection prompts, and includes R code chunks for prior elicitation and model updating.
    \item \texttt{4-evaluations}: Contains surey questions and anonymized student responses to the post-activity survey.
\end{enumerate}

An additional folder, \texttt{example-answers}, includes a representative completed student submission (\texttt{worksheet-group-01.qmd}) to assist instructors with assessment and rubric creation.

\bibliographystyle{plainnat} 
\bibliography{bibliography}  

\section*{Acknowledgements}

Both authors were affiliated with the Department of Statistics at the University of California, Irvine when working on this project.
Ricci gratefully acknowledges funding by the Hasso-Plattner-Institute Research Center in Machine Learning and Data Science at UCI. 
Dogucu was supported by the National Science Foundation Improving Undergraduate STEM Education program, with award number 2215879.

\end{document}